# The balance of knowledge flows

*Giovanni Abramo (corresponding author)*

Laboratory for Studies in Research Evaluation, Institute for System Analysis and Computer Science (IASI-CNR), National Research Council of Italy
giovanni.abramo@uniroma2.it

*Ciriaco Andrea D'Angelo*

Department of Engineering and Management, University of Rome "Tor Vergata" and Laboratory for Studies in Research Evaluation, Institute for System Analysis and Computer Science (IASI-CNR), Italy
dangelo@dii.uniroma2.it

*Massimiliano Carloni*

Clarivate Analytics, Customer Success Team, Italy
massimiliano.carloni@Clarivate.com

**Abstract**

In analogy to the technology balance of payments, in this paper we propose a possible way to set up a "balance of knowledge flows" (BKF), recording world flows of knowledge within the scientific community. Adopting a pure bibliometric approach, the "knowledge" traced in the BKF is that produced and exchanged by the scientific community by means of publications and relevant citations. A description of the theoretical foundation of such a tool is presented together with its empirical testing over the scientific production of four different countries. The BKF can be part of yearly reports of science and technology indicators, aimed at informing research policy.

# 1. Introduction

A country's balance of payments (BoP) records all economic transactions between the residents of the country and the rest of the world in a particular period. A country's BoP data may reveal, among others, its potential as a business partner for the rest of the world, and the performance of the country in international economic competition.

The technology balance of payments (TBP), a sub-division of the BoP, registers the commercial transactions related to international technology and know-how transfers. It consists of money paid or received for the use of patents, licenses, know-how, trademarks, patterns, designs, technical services (including technical assistance) and for industrial research and development (R&D) carried out or sold abroad. Although TBP data should be considered as only partial measures of international technology flows, they tend to reflect a country's ability to sell its technology abroad and its use of foreign technologies. On the relevance of the issue, almost half a century ago Mansfield claimed that "interchange of technical knowledge among nations importantly affects the pattern of world trade and influences economic growth rates" (Mansfield, 1974).

TBP data are provided, among others, for OECD member countries and seven non-member economies in a biannual publication by OECD, "Main Science and Technology Indicators".

In the realm of Science, currently we lack similarly constructed data on the world flows of knowledge. The aim of this work is to propose a possible way to set up a country's "balance of knowledge flows" (BKF) which records the outflows of knowledge produced in a country to other countries, and the inflows of knowledge produced by other countries in a country. A country's BKF shows a surplus when the difference between knowledge outflows and inflows is positive, a deficit when the opposite is true. In a sense, TBP embeds knowledge flows, but only knowledge embodied in the technologies exchanged among countries.[1] The BKF aims at recording the "exchange" of knowledge which has only partly or not yet been incorporated into technologies. The knowledge we are talking about is the one produced and exchanged by the scientific community. Because the scientists' principal goal is to produce new knowledge and diffuse it, they typically encode it in publications. The scientific knowledge most distinctive feature, which distinguishes it from goods and services, is that it is a public good. The immediate consequence is that its exchange cannot be traced through economic transactions. Other peculiar features of knowledge are: i) it is intangible, as its essence is "information"; ii) it is cumulative, as it builds on itself; and iii) it does not wear out physically, therefore can be used unlimited times without diminishing its substance. In other words, in simple terms, while a single product can be bought only by a single buyer, scientific knowledge encoded in publications can be appropriated by a multitude of scientists.

Three main questions need to be answered to design a BKF. First, how the scientific knowledge, object of the exchange, is measured. Second, how the country of production, the "made in" country, is identified. Third, how to certify that such knowledge has been used by the country of production itself and/or other countries. In answering the above mentioned questions, we also discuss the limits, assumptions, conventions, and caveats embedded in the construction of a BKF.

---

[1] Similarly, the BoP embeds technology flows, but only technologies embodied in the traded goods and services.

Scientific knowledge is encoded and spreads through scientific and technical literature, seminars, conferences, etc. Part of the new knowledge produced remains unavoidably tacit and can only be conveyed through personal communication between researchers. Bibliographic repertories, such as Web of Science (WoS) and Scopus, index publications of thousands of peer-reviewed scientific journals, and their coverage is continuously expanding. Either one of them can be a reliable source of the knowledge exchanged, as embodied in different publication types (articles, review articles, letters, conference proceedings, etc.). The limits are that: i) publications are not representative of all knowledge produced; ii) bibliographic repertories do not index all journals and consequently do not cover all publications.

Because of increasing international research collaboration, identifying the country of production of a publication may result not so straightforward. The U.S. National Science Foundation's annual report on Science and Engineering (S&E) Indicators provides an exhaustive compendium of bibliometric data on research collaboration. The latest edition (National Science Board, 2018) reports that in 2016, 64.7% of global S&E publications had multiple institutional addresses, compared with about 60.1% of such publications in 2006. The percentage of worldwide publications produced with international collaboration (i.e., by authors with institutional addresses from at least two countries) rose from 16.7% to 21.7% between 2006 and 2016. In particular, in the United States, 37.0% of publications were coauthored with researchers at institutions in other countries in 2016, compared with 25.2% in 2006.

In the case of internationally co-authored publications, a question needs to be answered: is it the nationality of the authors or the geographic location of the institutions they are affiliated to, which determines the country of production? Generally the second option is adopted (among others, Kim, 2006; Lewison & Cunningham, 1991; Schubert & Braun, 1990), because determining the nationality of authors is operationally formidable in large-scale analysis. Within this second option, various approaches for assigning an internationally authored publication to a country can be envisaged: i) to each country in the address list; ii) to one single country (based on the frequency that country, or the authors of that country, occur in the address list); iii) fractionalizing the publication by the number of countries, institutions or authors of a country.

The third question is how to identify users of knowledge. According to the Mertonian, normative conception of what citations signify and how they function (Kaplan, 1965; Merton, 1973), one can identify the authors of the citing publications as users of the knowledge encoded in the cited one. Since knowledge transfer cannot be observed directly (Jaffe, Trajtenberg, & Fogarty, 2000), one relies on proxy measures, notably citations. Citation linkages between articles is then assumed to imply a flow of knowledge from the cited to the citing entity (Mehta et al., 2010; Van Leeuwen & Tijssen, 2000). Stating that citations certify knowledge transfer or use does not imply that there are no exceptions, rather that it is the norm. Citations in fact are not always certification of real use and representative of all use. Uncitedness, undercitation, and overcitation may actually occur. Furthermore, citation-based analysis is unable to capture use outside the scientific system, such as that of practitioners (e.g. a physician applying a new pharmacological protocol after reading relevant literature), students, industry. It does not capture either tacit knowledge flows, such as those occurring in research collaborations.[2]

The ultimate goal of this paper is to show that measuring BKF is feasible, paralleling the measurement of TBP, and can be part of yearly reports of science and technology indicators. In the next section we review the scarce literature on the subject. Section 3 presents the data and method of analysis. Section 4 provides the results from the elaborations both at overall and at field level, with a final subsection devoted to the measurement of a country's comparative advantage in "exporting" and "importing" knowledge. Section 5 closes the work with our considerations on the relevance of the study.

## 2. Literature review

Knowledge flows across disciplines, more commonly termed as knowledge trade, has been largely studied in the literature (Cronin & Davenport, 1989, Cronin & Meho, 2008, Cronin & Pearson, 1990, Goldstone & Leydesdorff, 2006, Guerrero et al., Hessey & Willett, in press, Larivière et al., 2012, Lockett & McWilliams, 2005, Stigler, 1994). Early studies of knowledge flows were limited in their scope, focusing on either a single or a few disciplines (Cronin & Meho, 2008, Cronin & Pearson, 1990, Goldstone & Leydesdorff, 2006, Xhignesse & Osgood, 1967). Yan et al. (2013) were the first to provide a comprehensive overview of trends in knowledge flows across disciplines.

Very few studies instead delve into the geographic flows of scientific knowledge. This work intends to contribute to fill in the gap. The work lays its foundations on a recent study by the Abramo and D'Angelo (2018), investigating the domestic and transnational flows of scientific knowledge produced in Italy, how these vary across fields, and the comparative advantage of countries at benefiting from Italian research. The aim there was to identify the countries which most benefited from Italian research results, both in absolute and relative terms, in over 200 fields. Rabkin, Eisemon, Lafitte-Houssat, and McLean Rathgeber (1979) explored world visibility for four departments (botany, zoology, mathematics, and physics) of the universities of Nairobi (Kenya) and Ibadan (Nigeria), measured by citations in the Science Citation Index (SCI) for the years 1963-1977. They assessed the distribution of the author-country citing publications among five macro-regions (OECD, Eastern Europe, Africa, Latin America, and Asia), with specific attention to Great Britain, given its historic relations with Kenya and Nigeria. Their findings suggested high rates of domestic visibility for scientists in the two universities, mainly in botany and zoology, which are evidently locally oriented disciplines. However, not just for these two specific disciplines, the expectation was that in general, the main recipients of new knowledge produced by a country would be domestic scholars themselves. In fact the social links between the researchers of an individual country are on average stronger than those between researchers of different countries (Bozeman & Corley, 2004), as is confirmed by observations that rates of collaboration are higher domestically than internationally (Abramo, D'Angelo, & Murgia, 2013). At the level of the single field, Stegmann and Grohmann (2001) measured knowledge "export" and international visibility, through analysis of publication and citation data for the thirty journals listed in the Dermatology & Venereal Diseases category of the 1996 CD-ROM Journal Citation Reports (JCR), and in seven dermatology journals not listed in the 1996 JCR. Finally, Hassan and Haddawy (2013) mapped knowledge flows from the United States to other countries in the field of Energy over the years 1996-2009.

---

[2] We refer the reader to Abramo (2018) for a thorough discussion on the subject.

**3. Data and method**

As for the three main questions presented in the introduction to be answered to construct a country's BKF:
- We use a bibliometric approach assuming that new knowledge produced is measured by publications indexed in bibliographic repertories;
- By convention, we define a publication as "made in" a source country if at least 50% of the institutions authoring it belong to that country. To exemplify, if a publication is co-authored by three institutions, two of which located in country X, then we say that the publication is made in country X. If only one institution is located in country X, then the publication is not made in country X. The adopted convention is open to discussion;[3]
- We assume that knowledge flows are represented by "citations". In fact, when a publication is cited it has had an impact on scientific advancement because other scholars have drawn on it, more or less heavily, for the further advancement of science.

All limitations and assumptions typical of bibliometric analyses then apply.

We use the approach detailed in Abramo & D'Angelo (2018): when a publication is cited it has given rise to a "benefit". The number of "benefits" deriving from a publication equals the number of citations, and if the citing publication is co-authored by scholars from one or more foreign countries, the benefit has crossed an international boundary. In the case of a citing publication whose address list shows $n$ different countries, the same benefit (citation) is "gained" contemporaneously by $n$ countries, so we can say that it has given rise to $n$ equal "gains", one for each country listed in the affiliation list of the citing publication. A publication cited by $m$ other publications would give rise to $m$ benefits and $m \times n$ gains. Among the $n$ citing countries there could be also the country the cited publication is made in. In this case we define the relevant gain as "domestic".

For space reasons and ease of exposition we restrict our analysis to four countries, namely Israel, Italy, New Zealand (NZ), and The Netherlands (NL).[4] We assume that the world consists of the above four countries only. As a consequence, benefits and gains are only the ones generated by the knowledge flows within and among such countries, and the BKF is constructed for each country measuring the gains associated to the inflows and outflows of knowledge within the four-country world.

---

[3] It could be more correct to consider the number of authors rather than institutions, but developing appropriate algorithms would be much more complex. Alternative conventions, such as the affiliation of the corresponding author, or first and last authors in non-alphabetically ordered bylines, could be adopted as well; or any other alternative listed in the introduction.

[4] Apart from Italy, the authors' country, and the attention to include countries with and without geographic proximity, the choice of countries is otherwise merely random.

As a proxy of knowledge produced in such four countries, we consider WoS indexed publications (articles, reviews, letters, conference proceedings) over a five-year period, 2004-2008. The citing publications are observed up to 10/06/2017. The overall publications,[5] benefits and gains per country are shown in Table 1.

From the Israeli perspective, for example, Table 1 shows that in the five-year period under observation, Israeli institutions authored 76,509 WoS indexed publications. 58,725 of such publications (77%) present an address list where Israel appears at least in 50% of authors' affiliations. In turn, only slightly more than 60% of said publications are cited at 10/06/2017 in works authored by institutions of at least one of the four countries under analysis. This corresponds exactly to 35,546 publications, cited 238,025 times in the overall, with an average of 6.7 benefits per publication (238,025/35,546). On average, 1.04 countries (among the 4 under analysis), appropriate such benefits for a total amount of 247,128 gains (238,025*1.04), of which two-thirds domestic, which means they relate to citing publications authored by Israeli institutions.

This latest data has proved particularly significant compared to the other countries under analysis, for which the share of domestic gains is still higher, with the maximum amount registered for Italy accounting for 85.9%. Italy is the largest country in terms of number of publications, benefits and gains: compared to The Netherlands the amount of works containing at least 50% of domestic affiliations is definitely higher (81.8% vs 72.2%) whilst the average benefits per publication are lower by over one point (8.03 vs 9.09). New Zealand is the smallest country: publications, citations/benefits and gains are approximately half as many as those registered for Israel.

*Table 1: 2004-2008 scientific production and citations received at 10/06/2017 by each country in the dataset*

|  | Israel | Italy | New Zealand | The Netherlands |
|---|---|---|---|---|
| Publications | 76,509 | 325,504 | 39,740 | 181,339 |
| With at least 50% national addresses | 58,725 (76.8%) | 266,350 (81.8%) | 28,826 (72.5%) | 130,902 (72.2%) |
| Of which cited (a) | 35,546 (60.5%) | 159,484 (59.9%) | 17,162 (59.5%) | 81,099 (62.0%) |
| Total benefits (citations) (b) | 238,025 | 1,280,463 | 116,849 | 737,289 |
| Average benefits per cited "made in" publication (b/a) | 6.70 | 8.03 | 6.81 | 9.09 |
| Total gains (c) | 247,128 | 1,323,487 | 120,459 | 766,944 |
| Of which domestic | 164,688 (66.6%) | 1,136,810 (85.9%) | 85,490 (71.0%) | 571,236 (74.5%) |
| Average gains per benefit (c/b) | 1.04 | 1.03 | 1.03 | 1.04 |

---

[5] A publication is considered "made in" one of the four countries, if at least 50% of the institutions out of all "real" world institutions belong to the country, as it should be in the construction of a real BKF.

## 4. Results and analysis

In this section we analyze the knowledge flows among the four countries under observation, at both overall and field level. The aim is to show the application of our methodology, which can be extended to include all countries in the world.

### 4.1 The balance of knowledge flows at overall level

Table 2 shows the BKF of all four countries. For the sake of easier reading, we will be referring to the Israel case alone. As shown in Table 1, the "made in" Israel cited publications are 35,546. They generate a total of 247,128 gains, 164,688 of which appropriated by domestic institutions. The remaining 82,440 (as shown in row 2, column 2 of Table 2) is a prerogative of the other three countries under analysis. Such three countries publish 257,745 cited publications (cited foreign publications), generating a total of approximately 417,354 foreign gains, of which Israel appropriates 75,488[6] (18.1%). Israel final balance of gains is therefore positive (surplus) and equal to +6,952, given the imbalance between the generated foreign gains (82,440) and the earned gains (75,488).

*Table 2: BKF for the countries in the dataset (among parenthesis percentages out of total gains).*

| Country | Foreign gains generated (a) | Cited foreign publications | Foreign gains generated by foreign publications | Earned gains (b) | BKF (a-b) |
|---|---|---|---|---|---|
| Israel | 82,440 (33.4%) | 257,745 | 417,354 | 75,488 (18.1%) | +6,952 |
| Italy | 186,677 (14.1%) | 133,807 | 313,117 | 217,718 (69.5%) | -31,041 |
| New Zealand | 34,969 (29.0%) | 276,129 | 464,825 | 43,238 (9.3%) | -8,269 |
| The Netherlands | 195,708 (25.5%) | 212,192 | 304,086 | 163,350 (53.7%) | +32,358 |

Table 3 shows accurate data related to the flows at issue. Data on the main diagonal of the matrix illustrate the share of benefits generated by a country which remain within that same country (domestic gains).[7] 66.6% of gains generated by the scientific production of Israeli institutions remain within their country; Italy appropriates 20.5%; New Zealand 1.9% and The Netherlands 10.9%. In turn, the scientific production of Italian institutions generates knowledge outflows to the other three countries amounting to 3.3% to Israel, 1.5% to New Zealand, and 9.3% to The Netherlands.

Needless to say, the matrix may be read by column, in which case it will show an insight into the origin of the knowledge inflows of a country. For example, 4,777 (3.7%) out of 128,728 total gains earned by New Zealand, originate from works authored by Israeli institutions, 19,909 (15.5%) from Italian institutions, 18,552 (14.4%) from Dutch institutions and the remaining 66.4% from "in house" publications.

---

[6] 75,488 are the publications, authored by at least an Israeli institution, citing foreign publications.

[7] It is the ones' complement of the data in brackets in column 2, Table 2.

*Table 3: Overall import-export of knowledge (gains) between countries in the dataset (among parenthesis percentages out of total gains).*

|  | | Earning | | | |
|---|---|---|---|---|---|
| | Country | Israel | Italy | New Zealand | The Netherlands |
| Generating | Israel | 164,688 (66.6%) | 50,675 (20.5%) | 4,777 (1.9%) | 26,988 (10.9%) |
| | Italy | 43,819 (3.3%) | 1,136,810 (85.9%) | 19,909 (1.5%) | 122,949 (9.3%) |
| | New Zealand | 3,403 (2.8%) | 18,153 (15.1%) | 85,490 (71.0%) | 13,413 (11.1%) |
| | The Netherlands | 28,266 (3.7%) | 148,890 (19.4%) | 18,552 (2.4%) | 571,236 (74.5%) |

## 4.2 The BKF at field level

To conduct a stratification of the BKF at field level, we use the SCs assigned to the journals hosting the cited publications. We adopt a "full counting" approach, meaning that a publication published in a multi-category journal is fully assigned to each SC associated to the journal. The cited publications of our dataset are distributed over 217 SCs (out of a total 252 SCs), grouped in 13 scientific macro-areas.[8]

As an example, Table 4 shows the value of the Italian BKF in the twelve SCs falling in the area Earth and Space Sciences. In this area, the Italian 2004-2008 publications generate altogether 153,240 gains, 10% of which to other countries (namely 15,265). Vice versa, Italy totals 16,302 gains from publications by the other three countries. The overall balance is therefore negative and equal to 1,037 units. By analysing data related to the single SCs, it can be noted that half of them show a negative balance, accounting for a minimum of -1,107 in Environmental studies; and the remaining half show a positive balance, accounting for a maximum of +512 in Geochemistry & geophysics.

*Table 4: Italian BKF for the WoS subject categories falling in Earth and Space Sciences*

| Subject category | Foreign gains generated (a) | Earned gains (b) | BKF (a-b) |
|---|---|---|---|
| Geochemistry & geophysics | 1857 | 1345 | +512 |
| Geosciences, multidisciplinary | 3373 | 2942 | +431 |
| Water resources | 1554 | 1473 | +81 |
| Limnology | 397 | 318 | +79 |
| Mineralogy | 183 | 128 | +55 |
| Paleontology | 531 | 505 | +26 |
| Meteorology & atmospheric sciences | 1240 | 1273 | -33 |
| Geography, physical | 963 | 998 | -35 |
| Geology | 464 | 511 | -47 |
| Oceanography | 728 | 985 | -257 |
| Environmental sciences | 3317 | 4059 | -742 |
| Environmental studies | 658 | 1765 | -1107 |
| Total | 15265 | 16302 | -1037 |

When extending the analysis to all areas, SCs with a higher inclination to export (or import) new knowledge may be identified. Table 5 reports the first 10 SCs registering the lowest BKF values and the first 10 SCs registering the highest BKF values for Italy.

---

[8] Mathematics; Physics; Chemistry; Earth and Space Sciences; Biology; Biomedical Research; Clinical Medicine; Psychology; Engineering; Economics; Law, political and social sciences; Art and Humanities; Multidisciplinary Sciences. The macro-areas and the assignment of SCs to them were at some point defined by the Institute of Scientific Information (ISI), although no longer showing in current Clarivate Analytics bibliographic products. There is no multi-assignment of SCs to macro-areas.

SCs highly inclined to import, with a BKF value ranging between -2,922 (Chemistry, multidisciplinary) and -1,689 (Microbiology), are top of the list. Actually, the prevailing presence of life sciences SCs, with three SCs falling in Biology, three in Clinical Medicine, two in Chemistry and one in Biomedical Research, is quite evident.

In the lower section of the table, five SCs falling in Physics (with Astronomy & astrophysics on the top of the list, +5,280; Physics, particles & fields, +3,334; and Physics, nuclear, +995) and two SCs falling in Earth and Space Sciences stand out from among the SCs with a higher inclination to export.

The analysis of knowledge flows may also be carried out on pairs of countries in order to identify the SCs with the highest surplus or deficit. Table 6, for example, reports the analysis on the flows between Italy and The Netherlands for the SCs falling in Biomedical Research. The reported value of balances shows surplus and deficit by SC, from the Italian perspective. Overall, Italy is the country which imports most of the knowledge from The Netherlands (-4,869) mainly by virtue of the flows related to the SC Radiology, nuclear medicine & medical imaging. In reference to this SC, outflows from Italy to The Netherlands account for 58% of those observed in the opposite direction (namely 2,926 vs 5,064). Italy's BKF balance is negative in 10 more SCs, whilst it is positive only in Immunology (+264), Chemistry, medicinal (+169) and Anatomy & morphology (+32).

Table 7 shows the extension to all areas of the above mentioned analysis, with reference only to the bi-directional flows between Israel and New Zealand. Due to the size of these two countries, figures are definitely lower than those observed above. However, also in this case, SCs showing a greater spread between knowledge inflows and outflows from one country to another can be identified. Table 7 takes the Israel perspective in determining the BKF deficit or surplus. It is the other way around from the New Zealand perspective.

*Table 5: Subject categories (in Italy) with the highest and the lowest BKF*

| Subject category | Area* | Foreign gains generated (a) | Earned gains (b) | BKF (a-b) |
|---|---|---|---|---|
| Chemistry, multidisciplinary | 3 | 3576 | 6498 | -2922 |
| Biochemistry & molecular biology | 5 | 11254 | 13866 | -2612 |
| Genetics & heredity | 7 | 4249 | 6571 | -2322 |
| Radiology, nuclear medicine & medical imaging | 6 | 3673 | 5930 | -2257 |
| Psychiatry | 7 | 2916 | 5123 | -2207 |
| Cardiac & cardiovascular systems | 7 | 8788 | 10970 | -2182 |
| Chemistry, physical | 3 | 4455 | 6488 | -2033 |
| Ecology | 5 | 1848 | 3761 | -1913 |
| Materials science, multidisciplinary | 9 | 2306 | 4004 | -1698 |
| Microbiology | 5 | 2974 | 4663 | -1689 |
| … | - | - | - | + |
| Engineering, electrical & electronic | 9 | 3481 | 3132 | +349 |
| Physics, fluids & plasmas | 2 | 1288 | 936 | +352 |
| Physics, mathematical | 2 | 1334 | 948 | +386 |
| Chemistry, medicinal | 6 | 1434 | 1047 | +387 |
| Geosciences, multidisciplinary | 4 | 3373 | 2942 | +431 |
| Geochemistry & geophysics | 4 | 1857 | 1345 | +512 |
| Gastroenterology & hepatology | 7 | 4742 | 4054 | +688 |
| Physics, nuclear | 2 | 1592 | 597 | +995 |
| Physics, particles & fields | 2 | 5502 | 2168 | +3334 |
| Astronomy & astrophysics | 2 | 10998 | 5718 | +5280 |

*\* 1, Mathematics; 2, Physics; 3, Chemistry; 4, Earth and Space Sciences; 5, Biology; 6, Biomedical Research; 7, Clinical Medicine; 8, Psychology; 9, Engineering.*

*Table 6: The Italy - The Netherlands BKFs in the WoS subject categories of Biomedical Research*

| Subject category | Gains from Italy to NL (a) | Gains from NL to Italy (b) | BKF (a-b) |
|---|---|---|---|
| Radiology, nuclear medicine & medical imaging | 2926 | 5064 | -2138 |
| Oncology | 7933 | 9041 | -1108 |
| Infectious diseases | 1309 | 1738 | -429 |
| Toxicology | 967 | 1329 | -362 |
| Hematology | 5229 | 5582 | -353 |
| Pathology | 1018 | 1266 | -248 |
| Virology | 589 | 807 | -218 |
| Pharmacology & pharmacy | 3976 | 4189 | -213 |
| Medical laboratory technology | 526 | 632 | -106 |
| Medicine, research & experimental | 1926 | 2017 | -91 |
| Allergy | 726 | 794 | -68 |
| Anatomy & morphology | 82 | 50 | +32 |
| Chemistry, medicinal | 739 | 570 | +169 |
| Immunology | 5036 | 4772 | +264 |
| Total | 32982 | 37851 | -4869 |

*Table 7: The Israel - New Zealand BKFs in the bottom and top 10 WoS subject categories per Israel BKF deficit and surplus*

| Subject category | Area* | Gains from Israel to NZ | Gains from NZ to Israel | BKF |
|---|---|---|---|---|
| Toxicology | 6 | 22 | 82 | -60 |
| Endocrinology & metabolism | 7 | 88 | 144 | -56 |
| Psychology | 8 | 44 | 86 | -42 |
| Psychiatry | 7 | 108 | 138 | -30 |
| Food Science & technology | 5 | 76 | 98 | -22 |
| Nutrition & dietetics | 7 | 39 | 61 | -22 |
| Pharmacology & pharmacy | 6 | 136 | 155 | -19 |
| Psychology, clinical | 8 | 25 | 44 | -19 |
| Substance abuse | 7 | 1 | 20 | -19 |
| Parasitology | 7 | 33 | 51 | -18 |
| … | - | - | - | - |
| Plant sciences | 5 | 172 | 109 | +63 |
| Geography, physical | 4 | 84 | 17 | +67 |
| Clinical neurology | 7 | 141 | 69 | +72 |
| Microbiology | 5 | 151 | 79 | +72 |
| Management | 10 | 115 | 42 | +73 |
| Physics, multidisciplinary | 2 | 109 | 20 | +89 |
| Neurosciences | 7 | 288 | 195 | +93 |
| Chemistry, multidisciplinary | 3 | 132 | 26 | +106 |
| Biochemistry & molecular biology | 5 | 390 | 247 | +143 |
| Ecology | 5 | 378 | 143 | +235 |

*\* 1, Mathematics; 2, Physics; 3, Chemistry; 4, Earth and Space Sciences; 5, Biology; 6, Biomedical Research; 7, Clinical Medicine; 8, Psychology; 9, Engineering; 10, Economics.*

## 4.3 The knowledge flows specialization indexes

In this subsection, we present a way to measure the specialization indexes for outflows (export) and inflows (import) of knowledge by a given country. The relevant indicators, are the "knowledge outflows specialization index" (KOSI) and the "knowledge inflows specialization index" (KISI). They measure respectively a country's capacity to "export" knowledge to other countries, or to "import" knowledge from other countries, as compared to the rest of the world, across all research fields. In simple terms, they measure he extent to which a country's knowledge flows differ from those of the rest of the world or a comparison group of countries.

In operational terms, KOSI is calculated here applying the "revealed comparative advantage" (RCA) methodology and, in particular, the *Balassa index* (Balassa, 1979). The KOSI and KISI of country k in the SCj (respectively *KOSI$_{kj}$* and *KISI$_{kj}$*) are defined as:

$$KOSI_{kj} = 100 * \tanh ln \left\{ \frac{\left(G_{kj} / \sum_{i \neq j} G_{ki}\right)}{\sum_{z \neq k} G_{zj} / \sum_{z \neq k} \sum_{i \neq j} G_{zi}} \right\}$$

and

$$KISI_{kj} = 100 * \tanh ln \left\{ \frac{\left(G_{kj} / \sum_{i \neq j} G_{ki}\right)}{\sum_{z \neq k} G_{zj} / \sum_{z \neq k} \sum_{i \neq j} G_{zi}} \right\}$$

with $G_{kj}$ indicating the gains generated (KOSI) or earned (KISI) by country *k* in the SC*j*. Use of the logarithmic function centers the data around zero and the hyperbolic tangent multiplied by 100 limits the $KOSI_{kj}$ and $KISI_{kj}$ values to a range of +100 to -100. For any SC, the closer the value of the index to +100 the more the country is specialized in that SC in generating (appropriating) knowledge flows to (from) other countries. Vice versa, the closer the index approaches -100, the less the country is specialized in the SC. Values around 0 are labeled as "expected".

Table 8 shows the results of the application of this index for the outflows of knowledge, listing the ten SCs with the highest values of $KOSI_{kj}$, for each country. Similarly, Table 9 lists the ten SCs with the highest value of $KISI_{kj}$, for the inflows of knowledge in each country. With reference to Table 8, for Israel the list includes two SCs in Mathematics, as well as in Acoustic and Material science, with further SCs in Social Sciences (Anthropology; Area studies) and *humanities* (Literature; Archaeology; Dance, …) standing out from among the others. For Italy, the concentration of SCs in Physics is evident; as per New Zealand, with respect to the listed SCs six out of ten are in Biology and Medicine, whilst for The Netherlands the list does not show particular concentration per single areas. Obviously, based on the assumption of a four-country world, the outlined indexes do not constitute the actual specialization indexes of such countries, but only the results of a mere algebraic exercise. However, the analysis clearly shows the potential of this tool which, should it be applied to the real world, would illustrate the areas where a country is more likely specialized in exporting or importing new knowledge to and from other countries, exactly in the same way it is possible now to identify, through TBP data, the areas where a country is more likely specialized in exporting/importing new technology.

*Table 8: Subject categories with the highest knowledge outflows specialization indexes (in brackets) by country*

| Israel | Italy | New Zealand | The Netherlands |
|---|---|---|---|
| Acoustics (81.0) | Engineering, marine (91.0) | Materials science, textiles, paper & wood (99.7) | Engineering, ocean (88.9) |
| Anthropology (80.6) | Psychology, psychoanalysis (87.6) | Evolutionary biology (96.3) | Health care sciences & services (80.8) |
| Mathematics (78.6) | Physics, nuclear (87.5) | Substance abuse (94.3) | Geography (72.8) |
| Area studies (78.6) | Physics, particles & fields (85.8) | Medicine, legal (94.3) | Mycology (72.6) |
| Materials science, characterization & testing (75.9) | Astronomy & astrophysics (75.0) | Nursing (93.3) | Environmental studies (71.8) |
| Entomology (74.9) | Nuclear science & technology (70.3) | Geology (91.0) | Health policy & services (67.3) |
| Literature (73.2) | Mineralogy (63.8) | Toxicology (89.3) | Urban studies (63.5) |
| Archaeology (68.0) | Anatomy & morphology (58.4) | Law (88.8) | Religion (63.1) |
| Dance, theater, music, film and folklore (64.7) | Chemistry, medicinal (57.1) | Fisheries (88.6) | Information science & library science (61.8) |
| Mathematics, applied (63.5) | Telecommunications (56.3) | Construction & building technology (87.3) | Psychology, clinical (61.7) |

*Table 9: Subject categories with the highest knowledge inflows specialization indexes (in brackets) by country*

| Israel | Italy | New Zealand | The Netherlands |
|---|---|---|---|
| Psychology, psychoanalysis (95.5) | Mining & mineral processing (86.7) | Classics (88.3) | Engineering, marine (72.9) |
| Religion (82.2) | Materials science, textiles, paper & wood (63.5) | Oceanography (83.8) | History of social sciences (66.1) |
| Humanities, multidisciplinary (72.9) | Engineering, ocean (61.6) | Education, scientific disciplines (83.8) | Law (65.7) |
| Optics (72.7) | Environmental studies (51.1) | Biodiversity conservation (81.5) | Astronomy & astrophysics (61.1) |
| Physics, nuclear (71.8) | Classics (43.2) | Marine & freshwater biology (81.3) | Emergency medicine (54.5) |
| Communication (65.3) | Geography (42.4) | Ecology (81.2) | Political science (52.3) |
| Psychology, social (61.7) | Multidisciplinary sciences (37.7) | Architecture & art (80.5) | History (49.1) |
| Physics, atomic, molecular & chemical (61.3) | Radiology, nuclear medicine & medical imaging (37.6) | Soil science (80.3) | Remote sensing (45.7) |
| Archaeology (60.0) | Energy & fuels (37.4) | Geochemistry & geophysics (79.8) | Medieval & renaissance studies (45.6) |
| Physics, multidisciplinary (58.6) | Thermodynamics (37.1) | Engineering, geological (79.1) | Andrology (45.0) |

# 6. Conclusions

Data on commercial transactions related to international technology transfers are crucial for many reasons, first and foremost for the study of the determinants of macro-economic development (Mansfield, 1974) and of the total factor productivity of national systems (Mendi, 2007). To this purpose, TBP data are surveyed, updated and disseminated by national and international statistics bureaus all over the world. TBP embeds also knowledge flows, but only if "embodied" in the technologies formally exchanged among countries: it misses all other knowledge flows, especially those exchanged within the world scientific community.

Aim of this work was to fill in the gap, proposing a way to construct a country's balance of knowledge flows, which records the exchange (inflows and outflows) of knowledge proxied by citations to international scientific literature. The proposed bibliometric approach relies on the assumption that when a publication is cited it has had an impact on scientific advancement, because other scholars have drawn on it, more or less heavily, for the further advancement of science.

In this paper we have shown how the BKF can be constructed, at both overall and field level, and how the interchange of knowledge between countries can be analysed. We also showed how citation analysis can provide additional useful information to the policy maker, such as the share of domestic vs foreign gains generated by a country's research system, by field and as compared to other countries. A very high domestic share is expected in those fields where research is context specific or mainly oriented towards domestic needs. The knowledge outflows and inflows comparative advantage analysis can inform national research strategies, and be particularly pertinent concerning bilateral research collaboration agreements.

As it regards the BKF, i) the larger a country in terms of number of researchers and resources devoted to research; ii) the more productive its research system; and iii) the more scientifically advanced in terms of domestic stock and level of accumulated knowledge, as compared to other countries, the higher the chances that the new knowledge produced stems from domestic research rather than foreign. A country's BKF then should show a surplus in those fields where the country is research-leader (frontier research). The opposite occurs in those fields where it is a follower (gap-filling research).

Needless to underline that no conclusion may be reached on the countries analysed in this paper, as the world represented therein is limited to only four countries. Additionally, the conventions adopted, such as the way to identify the country of production of internationally authored publications, are open to discussion. Finally, all limitations and assumptions embedded in bibliometric analysis apply and caution is recommended in interpreting BKF. The measurement of the balance of knowledge flows for all countries, paralleling the measurement of technology balance of payments, can give rise to interesting comparative analyses, and be ultimately part of yearly reports of science and technology indicators.